\documentclass[a4paper]{jpconf}
\usepackage{graphicx}

\begin{document}
\title{Two dimensional frustrated magnets in high magnetic field}

\author{L. Seabra$^1$, 
N. Shannon$^1$, 
P. Sindzingre$^2$,
T. Momoi$^3$,\\
B. Schmidt$^4$
and P. Thalmeier$^4$}

\address{
 $^1$
  H. H. Wills Physics Laboratory, University of Bristol,
 Tyndall Ave, BS8-1TL, UK
\\  $^2$
 Laboratoire de Physique Th\'eorique de la Mati\`ere Condens\'ee, UMR
 7600 of CNRS,  
 \protect\mbox{Universit\'e P. et M. Curie, case 121,
4 Place Jussieu, 75252 Paris Cedex, France}
\\  $^3$
 Condensed Matter Theory Laboratory, RIKEN, Wako,
 Saitama 351-0198, Japan 
 \\  $^4$
 MPI-CPfS, No\"ethnitzer Str. 40, 01187 Dresden, Germany
}

\ead{luis.seabra@bristol.ac.uk}

\begin{abstract}
Frustrated magnets in high magnetic field have a long history of offering 
beautiful surprises to the patient investigator.   
Here we present the results of extensive classical Monte Carlo simulations of a variety of 
models of two dimensional magnets in magnetic field, together with complementary spin wave analysis.
Striking results include 
(i) a massively enhanced magnetocaloric effect in antiferromagnets bordering on ferromagnetic order, 
(ii) a route to an $m=1/3$ magnetization plateau on a {\it square} lattice, and 
(iii) a cascade of phase transitions in a simple model of AgNiO$_2$.

\end{abstract}

\section{Introduction}

It is precisely the feature that makes frustrated magnets so compelling --- the multitude of groundstates available --- that also makes them so difficult to study, especially if we take in account the extra contribution of an external magnetic field.  In this article we 
present a snapshot of ongoing work on several experimentally motivated models of two--dimensional frustrated magnets in high magnetic field.   Our main tool will be  classical Monte Carlo simulation, but where the standard local update Metropolis algorithm is augmented by a parallel tempering scheme \cite{Hukushima:1995kx}, which greatly improves the ergodicity of the method.   
Simulation results will also be compared with simple mean field theories and classical spinwave calculations.

One concept that will be used throughout this work is the magnetocaloric effect (MCE),
$
\theta_{MCE} = {\frac{\partial T}{\partial H}\mid}_S
$, 
the rate of change of temperature of the system with applied magnetic field at constant entropy.   This quantity is useful as a tool for identifying phase transitions in both simulation and experiment, and forms the basis for magnetic refrigeration.

 \section{The $J_1$-$J_2$ Heisenberg model on a square lattice}

The first model we consider is the Heisenberg model on a square lattice with additional 
\mbox{$2^{nd}$-neighbour} ($J_2$) bonds
\begin{equation}
\label{201}
H=J_1\sum_{\langle ij\rangle_1}\textbf{S}_i.\textbf{S}_j +
J_2\sum_{\langle ij\rangle_2}\textbf{S}_i.\textbf{S}_j 
- h \sum_i S^z_i .
\end{equation}

\begin{figure}[h]
\begin{minipage}{14pc}
\includegraphics[width=14pc]{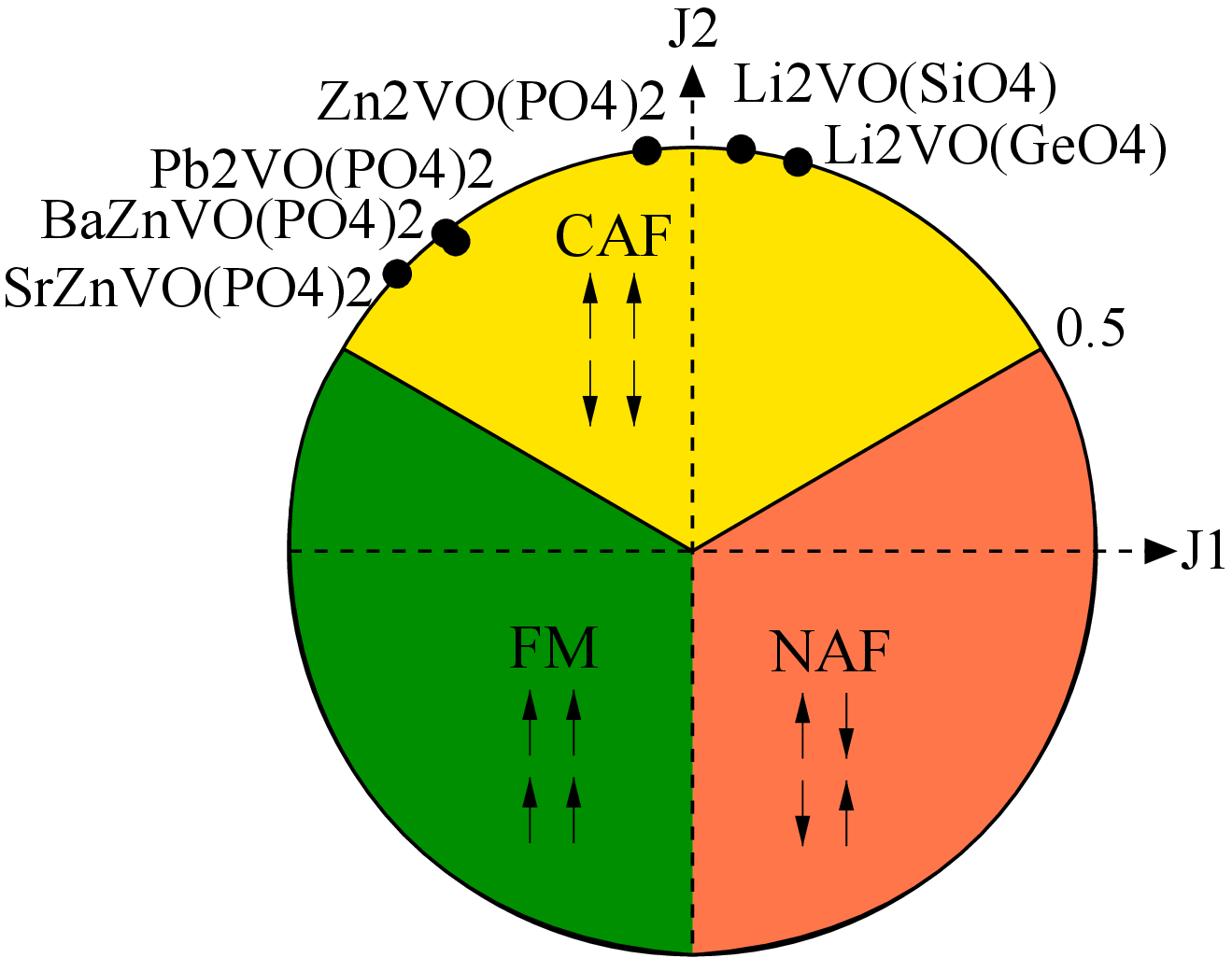}
\caption{\label{j1j2phase}(Colour online) Classical phase diagram of the $J_1$-$J_2$ Heisenberg model.   Parameters for quasi-2D vanadates are taken from~\cite{schmidt2007}.}
\end{minipage}\hspace{2pc}%
\begin{minipage}{20pc}
\begin{center}
\includegraphics[width=15pc]{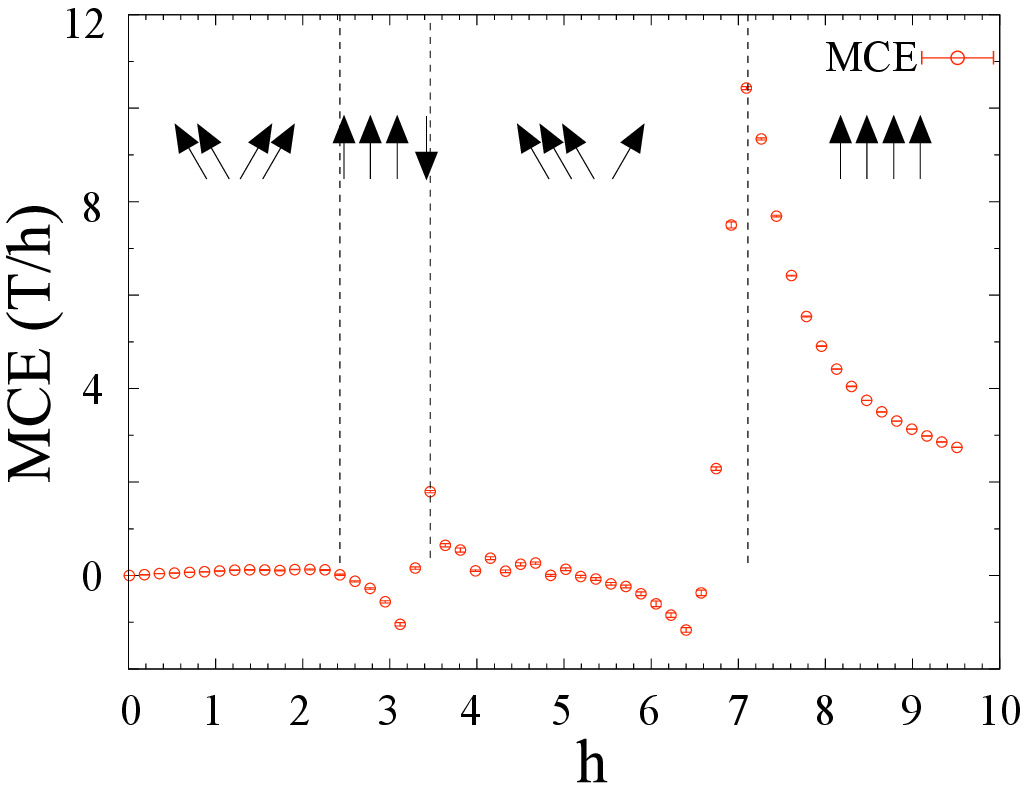}
\caption{\label{12plat} (Colour online) MCE of the classical $J_1$-$J_2$ Heisenberg model
for \protect\mbox{$J_1=1$, $J_2=0.5$} ($T=0.01$), showing successive phase transitions.   
$T$ and $h$ are measured in units of $J_1$.  The MCE is normalised to that of an ideal paramagnet.}
\end{center}
\end{minipage} 
\end{figure}

Despite its apparent simplicity, this is a very interesting model, and is thought to describe two new classes
of oxide compounds~\cite{Kaul:2004hc,kageyama2005}.    The classical phase diagram in zero field is shown in Fig.~\ref{j1j2phase}.
The quantum phase diagram and finite field properties of the spin-$1/2$ quantum model have been extensively investigated for both antiferromagnetic (AF)~\cite{misguich} and ferromagnetic (FM)~\cite{shannon2004ftp,shannon2006nos, schmidt2007,thalmeier2008} $J_1$.   New quantum phases are found in the highly frustrated regions where $|J_1|=2J_2$.   Here we explore the role of thermal fluctuations and finite magnetic field for these parameter sets.

Considering first AF $J_1>0$,  in Fig.~\ref{12plat} we present the MCE for $J_1=1$, $J_2=0.5$, as determined
by classical MC simulation for a cluster of $N = 24\times 24$ spins at $T=0.01$.   
Binned data from $\sim2 \times 10^6$  MC steps of simulation,  preceded by $\sim6 \times 10^6$ MC steps
of thermalization, were analysed using a jackknife procedure.   Analysis reveals four distinct phases, the most striking feature being  a $m=1/2$ plateau (c.f.~\cite{zhitomirsky2000fio}).   

\begin{figure}[h!]
   \includegraphics[height=5.1cm]{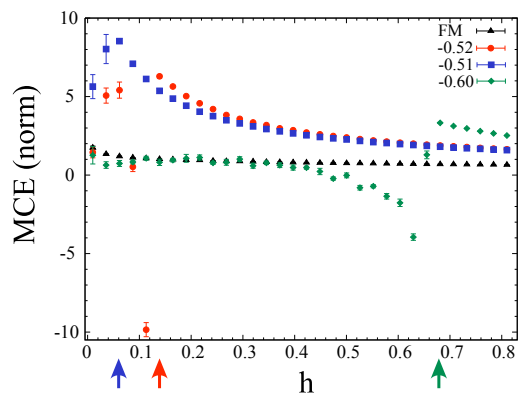}
\hspace{1pc}\begin{minipage}[b]{20pc}\caption{\label{emce} (Colour online).  MCE of the $J_1$-$J_2$ Heisenberg model on a square lattice for different ratios of $J_2/J_1$ ($J_1 <0$), showing a massive enhancement near the transition from collinear $(\pi,0)$ N\'eel AF to FM at $J_2/J_1=-0.5$.Arrows show rough values of the saturation field in each case. Results are for $24\times24$ spins, at $T=0.01$, normalised to the maximum value of the MCE in a nearest-neighbour Heisenberg AF.  $T$ and $h$ are measured in units of~$J_1$.}
\end{minipage}
\end{figure}

The MCE also shows interesting structure within the collinear AF phase for FM $J_1<0$, approaching the classical critical point 
$J_1=-2J_2$, as shown in Fig.~\ref{emce}.   Here the high degeneracy in the spin wave spectrum
shows up as a massive enhancement in the MCE at low fields, also discussed for quantum spins in~\cite{schmidt2007}. 

\section{The $J_1$-$J_2$-$J_3$ Heisenberg model on a square lattice}

A new series of layered Dion-Jacobi compounds containing spin-$1/2$ Cu($3d^9$) ions on a square lattice have recently been discovered~\cite{kageyama2005}.   These materials are believed to have FM nearest neighbour exchange $J_1$, but not all of their properties can be reconciled with a simple \mbox{$J_1$-$J_2$} model.   In particular (CuBr)Sr$_2$Nb$_3$O$_{10}$ exhibits an unexpected $m=1/3$ magnetization plateau~\cite{Tsujimoto:2007kx}.   This has motivated us to consider the 
quantum and classical properties of Eq.~\ref{201} with additional AF third neighbour interaction $J_3$.   

\begin{figure}[h!]
   \centering
   \includegraphics[height=5.3cm]{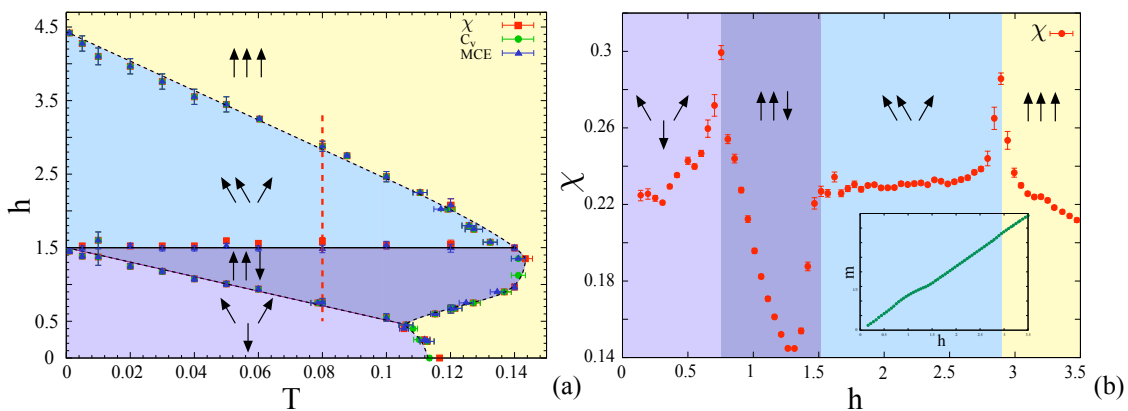}
   \caption{(Colour online.) Left Panel : Magnetic phase diagram for the classical $J_1$-$J_2$-$J_3$ Heisenberg model on a
   $24\times 24$ site square lattice, for $J_1=-1$, $J_2=1$, $J_3=0.5$.   Points are taken from simulation results for 
   magnetic susceptibility $\chi$, heat capacity $C_V$ and MCE.
   The vertical dashed line represents  a cut at $T=0.08$.  
   Right  Panel : the magnetic susceptibility on this cut. The inset displays the average magnetization revealing   the $m=1/3$ plateau.   
   $T$ and $h$ are measured in units of $J_1$.}
   \label{j3pic1}
 \end{figure}

So far as classical MC simulations are concerned, our main results are summarised in Fig.~\ref{j3pic1}. 
A clear $m=1/3$ plateau is found for $J_1=-1$, $J_2=1.0$, $J_3=0.5$.    The classical $h$-$T$ phase diagram clearly has much in common with that for the nearest-neighbour Heisenberg AF on a triangular lattice~\cite{miyashita}.  
The $m=1/3$ plateau is also observed in exact diagonalization calculations for the equivalent 
spin-$1/2$  $J_1$-$J_2$-$J_3$ model~\cite{philippeHFM}.   These results clearly suggest that this model deserves further 
study, both as a ``toy model'' for (CuBr)Sr$_2$Nb$_3$O$_{10}$ and as an interesting problem in its own right.

%
 
\section{The easy-axis $J_1$-$J_2$ Heisenberg model on a triangular lattice}

The metallic layered silver nickelate 2H-AgNiO$_2$ has been argued~\cite{wawrzynska2007cms} to 
offer realisation of a spin-$1$ easy-axis AF on a {\it triangular} lattice with competing AF $J_2$ interactions   
 \begin{equation}
\label{J1J2easyaxisHamiltonian}
H=J_1\sum_{\langle ij\rangle_1}\textbf{S}_i.\textbf{S}_j +
J_2\sum_{\langle ij\rangle_2}\textbf{S}_i.\textbf{S}_j 
- D \sum_i (S^z_i)^2
- h \sum_i S^z_i
\end{equation} 
We have investigated the high field properties of Eq.~\ref{J1J2easyaxisHamiltonian} with ratio of parameters $J_1=1.0$, $J_2=0.25$ and $D=0.06$ taken from preliminary fits to experiment~\cite{radu-unpub}.  For these parameters, the classical ground state of Eq.~\ref{J1J2easyaxisHamiltonian} is a two-sublattice collinear N\'eel state with ${\bf q} = (0, \pi)$.   A natural expectation would be that, in applied magnetic field, this would undergo a 1$^{st}$ order spin flop transition into a canted two-sublattice state with the same wave number.   However our results support a very different scenario.   

\begin{figure}[h!]
   \centering
   \includegraphics[height=4.5cm]{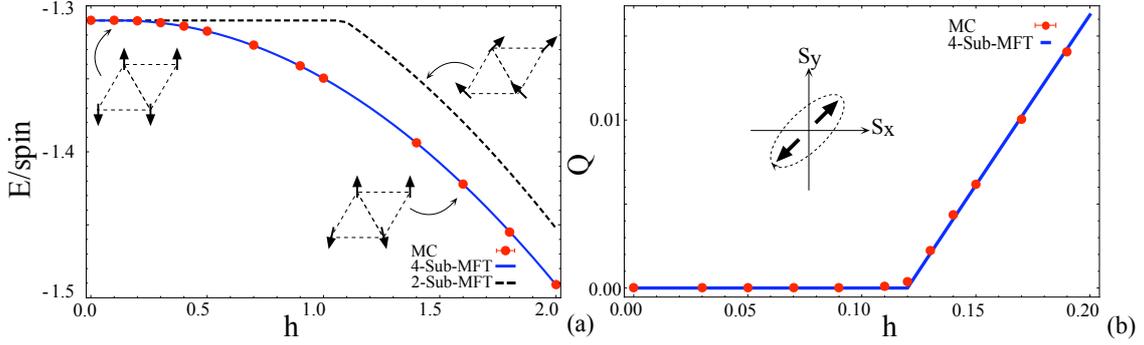}
   \caption{(Colour online) Left Panel : Ground state energies of Eq.~\ref{J1J2easyaxisHamiltonian} in 2-sublattice (dashed line) and 4-sublattice mean field approximations (blue line), together with results of classical MC simulation at $T=1\times10^{-4}$
(red points).   Simulation results follow the 4-sublattice solution for $h \ge 0.12$, avoiding the 2-sublattice
spin-flop for $h=1.1$.  Right Panel :  The quadrupole moment Q, showing a continuous transition into a state with transverse order.}
   \label{agnio2}
 \end{figure}

In Fig.~\ref{agnio2}-(a) we show results of classical MC simulations of  Eq.~\ref{J1J2easyaxisHamiltonian}, 
compared with a mean field theory (MFT) of the 2-sublattice spin-flop transition, and a 4-sublattice MFT.
The 2-sublattice theory predicts a first order spin flop transition at $h=1.1$.   However a spin wave with
${\bf q} = (\pi,\frac{\sqrt 3\pi}{3})$ becomes soft at the much lower field of $h=0.12$, precipitating a 2$^{nd}$ order transition
into a 4-sublattice ``supersolid'' state with both transverse order and broken translational symmetry.  
We can identify this transition using the quadrupole moment 
\begin{eqnarray}
Q=\sqrt{ Q_{x^2-y^2}^2 + Q_{xy}^2 } 
\qquad  
Q_{x^2-y^2} = \frac{1}{N} \sum_i (S^x_i)^2 -  (S^y_i)^2 
\qquad 
Q_{xy} = \frac{1}{N} \sum_i 2 S_i^x S^y_i
\end{eqnarray}
shown in Fig.~\ref{agnio2}-(b).  
We have checked in interacting spin wave theory that this ``supersolid'' scenario also holds for quantum spins.  
Our preliminary results for the global $h$-$T$ phase diagram also suggest that this supersolid state is superseded by cascade of phase transitions at higher field.   These will be reported elsewhere.
  
\section*{Acknowledgements}

We are pleased to acknowledge helpful conversations with Pierre Adroguer, Tony Carrington, Amalia and Radu Coldea, Hiroshi Kageyama, 
Karlo Penc, Yuki Motome, Daisuke Tahara, and Mike Zhitomirsky.

\bibliographystyle{iopart-num}
\section*{References}

\providecommand{\newblock}{}

\end{document}